\documentclass[runningheads]{llncs}
\usepackage{times}

\usepackage{cite}
\usepackage{hyperref}
\usepackage{bibnames}
\usepackage{graphicx}
\usepackage[pdftex,dvipsnames]{xcolor}  
\usepackage{amsmath,amssymb}
\usepackage{listings}
\usepackage{multirow}
\usepackage{longtable}
\usepackage{arydshln}
\usepackage{enumitem}
\usepackage{amsfonts}
\usepackage{multicol}

\newcommand{\Nat}{\mathbb{N}}
\newcommand{\Integer}{\mathbb{Z}}
\newcommand{\Real}{\mathbb{R}}

\usepackage{algorithm}
\usepackage[noend]{algpseudocode}

\makeatletter
\def\BState{\State\hskip-\ALG@thistlm}
\makeatother

\usepackage{xargs}                      
\usepackage[colorinlistoftodos,prependcaption,textsize=tiny]{todonotes}

\newcommandx{\unsure}[2][1=]{\todo[linecolor=red,backgroundcolor=red!25,bordercolor=red,#1]{#2}}
\newcommandx{\change}[2][1=]{\todo[linecolor=blue,backgroundcolor=blue!25,bordercolor=blue,#1]{#2}}
\newcommandx{\info}[2][1=]{\todo[linecolor=OliveGreen,backgroundcolor=OliveGreen!25,bordercolor=OliveGreen,#1]{#2}}
\newcommandx{\improvement}[2][1=]{\todo[linecolor=Plum,backgroundcolor=Plum!25,bordercolor=Plum,#1]{#2}}
\newcommandx{\thiswillnotshow}[2][1=]{\todo[disable,#1]{#2}}

\newcommand{\ProbModel}{Prob-solvable}
\newcommand{\Aligator}{{\tt Aligator}}

\newcommand{\Mom}[3]{Mom_{#1}[#2,#3]}

\newcommand{\nin}{\not\in}

\begin{document}

\title{Automatic Generation of Moment-Based \\ Invariants for Prob-Solvable Loops}
\titlerunning{ }
\authorrunning{ }
\author{ }
%
\author{Ezio Bartocci \and
Laura Kov{\'{a}}cs \and
Miroslav Stankovi\v{c}}
%
%
\institute{TU Wien, Austria
}

%
\maketitle              
%


\begin{abstract}
One of the main challenges in the analysis of probabilistic programs is to compute 
invariant properties that summarise loop behaviours. Automation
of invariant generation is still at its infancy and most of the times
targets only expected values of the program variables, which is insufficient 
to recover the full probabilistic program behaviour.  
We present a method to automatically generate {\it moment-based invariants} of
a subclass of probabilistic programs, called \ProbModel{} loops, with
polynomial assignments over random variables and parametrised
distributions. We combine methods from symbolic summation and
statistics to derive invariants as valid properties over higher-order
moments, such as expected values or variances, of program variables.  We
successfully evaluated our work on several examples where full
automation for computing higher-order moments and invariants over
program variables was not yet possible. 
\end{abstract}

\section{Introduction}
\label{sec:intro}

\begin{figure}[t]
  \label{fig:pp1}
  \centering
    \includegraphics[width=\textwidth]{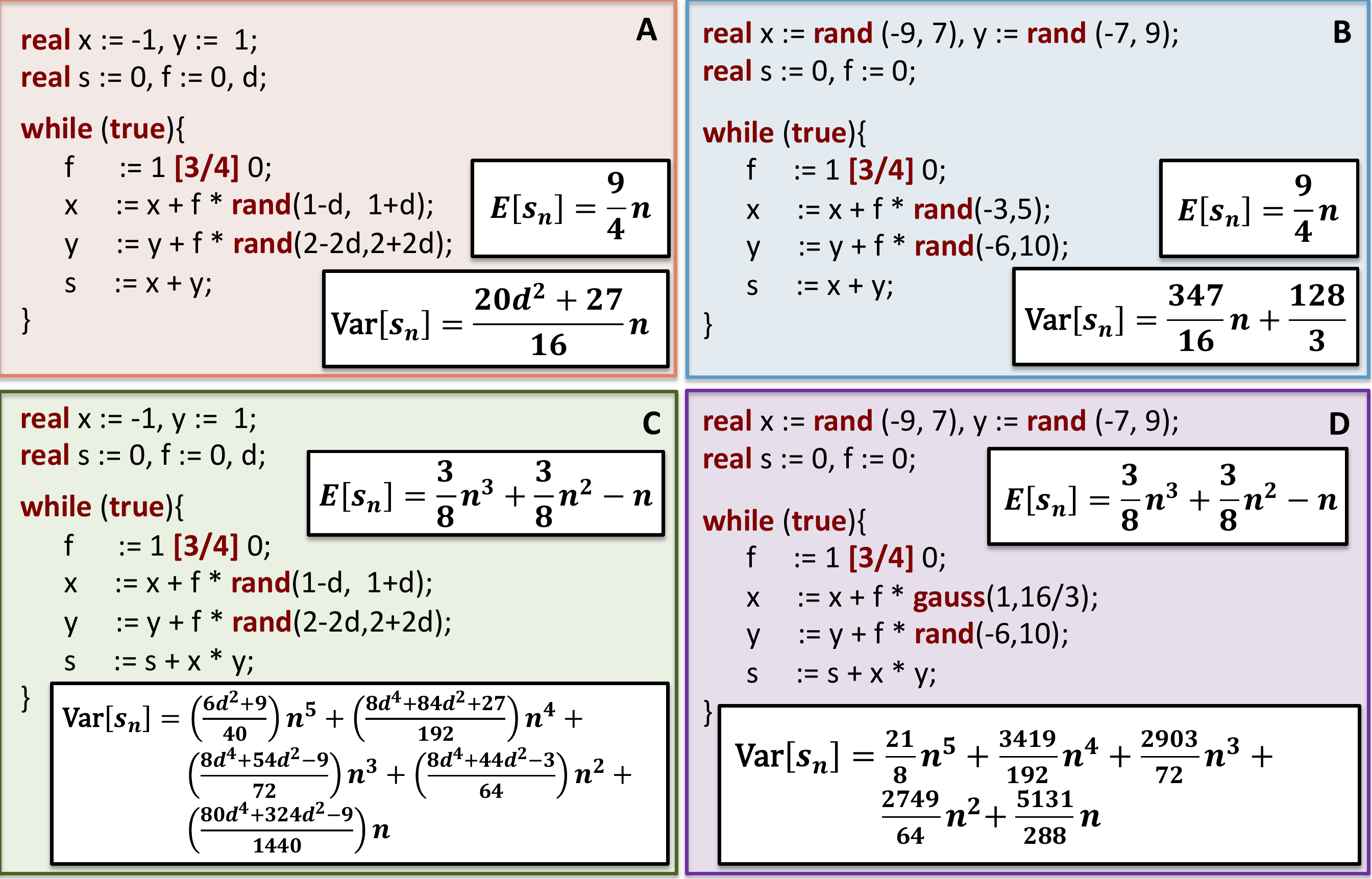}
    \vspace{-3ex}
      \caption{Examples of four \ProbModel{} loops.  ${\tt f:=1 [3/4] 0}$ is 
      a statement that assigns to $\tt f$ the value $1$ with probability $\frac{3}{4}$ and 
      the value $0$ with probability $1- \frac{3}{4} = \frac{1}{4}$. The function ${\tt rand(a,b)}$ 
      samples a random 
        number from a uniform distribution with support in the real interval $[a,b]$ and the 
        function  ${\tt gauss(\mu,\sigma^2)}$ samples a random 
        number from a normal distribution with mean $\mu$ and variance $\sigma^2$.
         For each loop, we provide 
        the moment-based invariants for
        the first ($E[]$) and second moments ($Var[]$)
        of $s$ computed using our approach, where $n$ denotes
        the loop counter. }
        \vspace{-4ex}
\end{figure}

Probabilistic programs (PPs), originally employed in
cryptographic/privacy protocols and randomised algorithms,
are now gaining momentum due to the 
several emerging applications in the areas of machine learning and
AI~\cite{Ghahramani15}.  
%
One of the main problems that arise from introducing randomness 
into the program is that we can no longer view variables as single 
values; we must think about them as distributions. 
Existing approaches, see e.g.~\cite{McIverM05,Katoen2010,Chakarov2014}
usually take into consideration only  expected values, or upper 
and lower bounds over program variables.  As argued
by~\cite{NoviInverardi:2006}, 
such information is however insufficient to characterize the full
value distributions of variables; (co-)variances and other
higher-order moments of variables are also needed.
Consider for example the PPs of Fig.~\ref{fig:pp1}(A) and Fig.~\ref{fig:pp1}(B):  the expected value of 
variable $s$ at each loop iteration is the same in both 
PPs, while the variance of the value distribution of $s$ differs in
general (a similar behaviour is also
exploited by Fig.~\ref{fig:pp1}(C)-(D)).
Thus, Fig.~\ref{fig:pp1}(A) and Fig.~\ref{fig:pp1}(B) do not have the same
invariants over higher-order moments; yet, current approaches would
fail identifying such differences  and only compute
expected values of variables. 
%

One of the main challenges in analysing PPs and computing their
higher-order moments comes with the presence of loops and the burden
of computing so-called \emph{quantitative
  invariants}~\cite{Katoen2010}. Quantitative invariants are
properties that are true before and after each loop iteration.
Weakest 
pre-expectations~\cite{McIverM05,Katoen2010} can be used to compute quantitative
invariants. This approach, supported for example in ~\textsc{Prinsys}~\cite{GretzKM13}, consists
in annotating a loop with a template invariant and then solve
constraints over the unknown coefficients of the template. 
Other methods~\cite{Barthe2016,Kura19} use martingales that are expressions over 
program variables whose expectations remain invariant.
The aforementioned approaches {are however not fully automatic since 
they require user guidance for providing templates and hints}. In
addition, they {are limited to invariants over only expected values}: 
with the exception of~\cite{Kura19}, they do not compute higher-order
moments describing the 
distribution generated by the PP (see Section~\ref{sec:related} for
more details). 

In this paper we introduce a {\it fully automated
approach} to compute invariant properties over {\it higher-order moments} of
so-called {\it \ProbModel{} loops}, to stand for {\it probabilistic
  P-solvable loops}. \ProbModel{} loops are PPs that extend the 
  imperative P-solvable loops described in~\cite{Kovacs2008} with probabilistic assignments over random
variables and parametrised distributions. As such, variable updates 
are expressed by random polynomial, and not only affine, updates (see
Section~\ref{sec:programmingmodel}).  Each program in
Fig.~\ref{fig:pp1} is \ProbModel{}; moreover, Fig.~\ref{fig:pp1}(C)-(D)
involve nonlinear 
updates over $s$. 

Our work uses statistical properties to eliminate probabilistic choices
and turn random updates into
recurrence relations over higher-order moments
of program variables. We show that higher-order moments of
\ProbModel{} loops can be described by C-finite recurrences
(Theorem~\ref{thm:line7Alg}). We further solve such recurrences
to derive {\it moment-based invariants} of
\ProbModel{} loops (Section~\ref{sec:algorithm}). A moment-based invariant is a property that holds at 
arbitrary loop iterations (hence, invariants), expressing closed form
solutions of higher-order moments of program variables. To the best of
our knowledge, no other method is able to derive higher-order moments
of PPs in a fully automated approach. Our work hence allows to
replace, for example, the required human guidance of~\cite{GretzKM13,KwiatkowskaNP11} for
\ProbModel{} loops.  Unlike existing works, we support 
PPs with parametrised distributions (e.g., in Fig.~\ref{fig:pp1}(A)):
instead of taking concrete instances of a given parametrised distribution,
we automatically infer
invariants of the entire class of PPs characterised by the
considered parametrised distribution. 

Our approach is both sound and terminating: given a
\ProbModel{} loops and an integer $k\geq 1$, we automatically
infer the moment-based invariants over the $k$th moments of our input
loop (see Section~\ref{sec:algorithm}). Unlike the approach of~\cite{Kovacs2008} for
deriving polynomial invariants of non-probabilistic (P-solvable)
loops, our work only computes closed form expressions over
higher-order moments and does not employ Gr\"obner basis computation
to eliminate loop counters from the derived closed forms. As such, our
moment-based invariants are not restrictive to polynomial properties 
but are linear combinations of polynomial expressions and exponential
sequences over the loop counter.  Moreover, \ProbModel{} are more
expressive than P-solvable loops as they are not restricted to
deterministic updates but allow random assignments over variables.


\noindent\paragraph{\bf Contributions.} Our main contributions are: (1) we introduce the class of \ProbModel{} loops with
probabilistic assignments over random variables and distributions
(Section~\ref{sec:programmingmodel}); 
(2) we show that \ProbModel{} loops can be modelled as C-finite recurrences
over higher-order moments of variables (Theorem~\ref{thm:line7Alg}); 
(3) we provide a fully automated approach that derives moment-based
invariants over arbitrary higher-order moments of \ProbModel{} loops (Algorithm~\ref{algo:InvGen});
(4) we implemented our work as an extension of the \Aligator{}
package~\cite{Aligator18} and evaluated over several challenging PPs (Section~\ref{sec:implementation}).



\section{Preliminaries}
\label{sec:preliminaries}

We recall  basic mathematical properties about  
recurrences and higher-order moments of variable values --  for more details see~\cite{Kauers11,Lin92}.
Throughout this paper, let $\Nat,\Integer, \Real$ denote 
the set of natural, integer and real numbers. We reserve capital
letters to denote abstract random variables, e.g. $X,Y,\ldots$, and use small
letters to denote program variables, e.g. $x,y,\ldots$, all  possibly with
indices.

\subsection{C-Finite Recurrences}
While sequences and recurrences are defined over arbitrary fields of characteristic
zero, in our work we only focus over sequences/recurrences over $\Real$.

\begin{definition}[Sequence]
A univariate \emph{sequence} in $\Real$ is a function
$f:\Integer\rightarrow\Real$. 
A \emph{recurrence} for a sequence $f(n)$ is  %
\[
f(n+r)=R(f(n),f(n+1),\dots,f(n+r-1),n),\qquad\qquad  \text{with } n \in \Nat,
\]
for some function $R:\Real^{r+1}\rightarrow\Real$, where $r  \in \Nat$ is
called the \emph{order} of the recurrence.
\end{definition}

For simplicity, 
we denote by $f(n)$ both the recurrence of $f(n)$ as well as the
recurrence equation $f(n)=0$. 
When solving the recurrence $f(n)$, one is interested in computing
a  {\it closed form} solution of $f(n)$, expressing the value of $f(n)$ as a function of $n$ for any $n\in\Nat$.
In our work we only consider the class of {\it linear recurrences with
  constant coefficients}, also called {\it C-finite recurrences}. 

\begin{definition}[C-finite recurrences] 
  A \emph{C-finite recurrence} $f(n)$ satisfies the linear homogeneous
  recurrence with constant coefficients: 
\begin{equation}\label{eq:CFinDef}
f(n+r)=a_0 f(n)+a_1 f(n+1) + \ldots + a_{r-1} f(n+r-1),\qquad\qquad
\text{with }r,n\in \Nat, 
\end{equation}
where $r$ is the \emph{order} of the recurrence, and
$a_0,\dots,a_{r-1}\in\Real$ are constants with $a_0 \neq 0$.
\end{definition}
%
%
An example of  a C-finite recurrence is the recurrence of
Fibonacci numbers satisfying the recurrence $f(n+2)=f(n+1)+f(n)$,
with initial values $f(0)=0$ and $f(1)=1$.  Unlike arbitrary recurrences, closed forms of  C-finite recurrences
 $f(n)$ always exist~\cite{Kauers11} and are defined as:
\begin{equation}\label{eq:c-finite-solving}
f(n)=P_1(n)\theta_1^n+\cdots+P_s(n)\theta_s^n,
\end{equation}
where $\theta_1,\ldots,\theta_s \in {\Real}$ are the distinct
roots of the characteristic polynomial of $f(n)$ and $P_i(n)$ are 
polynomials in $n$.  Closed forms of C-finite recurrences are  called \emph{C-finite expressions}.
We note that, while the C-finite
recurrence~\eqref{eq:CFinDef} is homogeneous, inhomogeneous C-finite
recurrences can always be translated into homogeneous ones, as the
inhomogeneous part of a C-finite recurrence is a C-finite
expression.

In our work, we focus on the analysis of \ProbModel{} loops and 
consider loop variables
$x$
as sequences $x(n)$, where $n\in\Nat$ denotes the loop iteration
counter.
Thus,  $x(n)$ gives the value of the program variable $x$ at iteration $n$.

\subsection{Expected Values and Moments of Random Variables}

Here we introduce the relevant notions from statistics that our work
relies upon.

\begin{definition}[Probability space]
	A \emph{probability space} is a triple $(\Omega, F, P)$ consisting of
	a sample space $\Omega$ denoting the set of outcomes, where $\Omega \not=\emptyset$,
	a $\sigma$-algebra $F$ with  $F \subset 2^\Omega$, denoting a set of events,  
	a probability measure $P: F\rightarrow [0,1]$ s.t. $P(\Omega)=1$.
\end{definition}

We now define random variables and their higher-order
moments. 

\begin{definition}[Random variable]
	A \emph{random variable} $X: \Omega \rightarrow \Real$ is
        a measurable function from a set $\Omega$ of 
	possible outcomes to $\Real$.
\end{definition}

In the context of our \ProbModel{} loops, for each loop variable $x$, 
we consider elements of its corresponding sequence $x(n)$ 
to be random variables.
When working with a random variable $X$, one is in general interested in
expected values and other moments of $X$.

\begin{definition}[Expected value]
An \emph{expected value of a random variable $X$} defined on a probability space $(\Omega, F, P)$ is 
the Lebesgue integral: $E[X] = \int_\Omega X\cdot dP.$
In the special case \emph{when  $\Omega$ is discrete}, that is the
outcomes are ${X_1,\dots X_N}$ with corresponding 
probabilities ${p_1,\dots p_N}$, we have $E[X] = \sum_{i = 1}^N X_i\cdot p_i.$ 
The expected value of $X$ is often also referred to as the
\emph{mean} or $\mu$ of $X$.
\end{definition}

For program variables $x$ of \ProbModel{} loops, our work computes
the expected values of the corresponding sequences $x(n)$ but also 
higher-order and mixed moments.

\begin{definition}[Higher-Order Moments]
Let $X$ 
be a random variable, $c\in\Real$ and $k\in\Nat$. We write
$\Mom{k}{c}{X}$ to denote the \emph{$k$th moment about $c$ of $X$},
which is defined as:  
\begin{equation}\label{eq:kthMoments}
  \Mom{k}{c}{X}=E[(X-c)^k]
 \end{equation}
\end{definition}
In this paper we will be almost solely interested in moments about $0$ (called \emph{raw moments}) 
and about the mean $E[X]$ (called \emph{central moments}). We note
though that 
we can move to moments of $X$ with different centers using Proposition~\ref{prop:transformation}.

\begin{proposition}[Transformation of center]
  \label{prop:transformation}
  Let $X$ 
be a random variable, $c,d\in\Real$ and $k\in\Nat$.  The \emph{$k$th moment about $d$ of $X$},
can be calculated from moments about $c$ of $X$ by: 
$\displaystyle E\left[(X-d)^{k}\right]=\sum _{i=0}^{k}{k \choose i}E\left[(X-c)^{i}\right](c-d)^{k-i}.$
\end{proposition}

Similarly to higher-order moments, we also consider \emph{mixed moments}, 
that is $E[X\cdot Y]$, where $X$ and $Y$ are random variables. 
For arbitrary random variables $X$ and $Y$, we have the following basic
properties about their expected values and other moments: 
\begin{itemize}
	\item $E[c] = c$ for a constant $c\in\Real$,
	\item expected value is linear, $E[X+Y] = E[X] + E[Y]$ and $E[c\cdot X] = c\cdot E[X]$,
	\item expected value is not multiplicative, in general $ E[X\cdot Y] \not= E[X]\cdot E[Y]$
	\item expected value is multiplicative for independent random variables.
        \end{itemize}
As a consequence of the above, expected values of monomials over
arbitrary random variables, e.g. $E[X\cdot Y^2]$, cannot be in general further simplified.

The moments of a random variable~$X$ with bounded support fully characterise its value
distribution. While computing all moments of~$X$ is generally
very hard, knowing only a few moments of~$X$ gives  
useful information about its value distributions. 
The most common 
moments are
variance, covariance, skewness, as defined below. 



\begin{definition}[Common moments]\label{def:common_Moments}
	\emph{Variance} measures how spread the distribution is and is
        defined as the second central moment: $Var[X] =
        \Mom{2}{E[X]}{X}$.
        
	\emph{Covariance} is a mixed moment measuring variability of two distributions
        and is defined as: $Cov[X,Y] = E[ (X - E[X]) \cdot (Y-E[Y]) ]$.

        \emph{Skewness} measures asymmetry of the distribution and is
        defined as the normalised third central moment: 
        $Skew[X]=\frac{\Mom{3}{E[X]}{X}}{(Var[X])^{3/2}}$.
        
\end{definition}

Basic results about variance and covariance state: 
$Cov[X,X] = Var[X]$, 
$Var[X] = E[X^2] -(E[X]^2)$, and 
$Cov[X,Y] = E[X\cdot Y] - E[X]\cdot E[Y]$.

\begin{definition}[Moment-Generating Function (MGF)]
	A moment generating function of a random variable $X$ is given
        by:  
	\begin{equation}\label{eq:MGF}
          M_X(t) = E[e^{tX}], \quad \text{with}\ t \in\Real
         \end{equation}
	whenever this expectation exists.
\end{definition}
Moment-generating functions, as the name suggests, can be used to
compute higher-order moments of a random variable $X$. 
If we take the $k$th derivative of the moment-generating function of $X$, evaluated at $0$, we get the $k$th moment 
about $0$ of $X$, that is $\Mom{k}{0}{X}$\footnote{due to the series
  expansion $e^{tX} = 1 + tE[X] + \frac{t^2E[X^2]}{2!} +
  \frac{t^3E[X^3]}{3!} + \dots$ and derivative {w.r.t.}~$t$}. For many standard
distributions, including Bernoulli, uniform and normal distributions, the moment-generating function 
exists and gives us a way to compute the moments for random variables drawing from these distributions. Thanks to 
these properties, we can use common distributions
in our \ProbModel{} programs.



\section{Programming Model: \ProbModel{} Programs}
\label{sec:programmingmodel}

We now introduce our programming model of  {\it
  \ProbModel{} programs}, to stand for {\it probabilistic P-solvable programs}.
P-solvable programs~\cite{Kovacs2008}  are
non-deterministic loops whose
behaviour can be expressed by a system of C-finite recurrences over
program variables. 
\ProbModel{} programs extend
P-solvable programs
by allowing probabilistic assignments over random variables and
distributions.

\vspace{-2mm}
\noindent\paragraph{\bf \ProbModel{} Loops.} 
Let $m\in\Nat$ and $x_1,\ldots x_m$ denote real-valued program
variables. 
We define \ProbModel{} loops with $x_1,\ldots x_m$
variables as programs of the form:
\begin{equation}\label{eq:ProbModel}
  I; \texttt{while(true)} \{U\}, \qquad\qquad\text{where:}
\end{equation}

\begin{itemize}
\item $I$ is a sequence of initial assignments over $x_1,\ldots, x_m$.
  That is, $I$ is an assignments sequence $x_1 := c_1; x_2 :=
c_2; \dots x_m := c_m$, with $c_i\in\Real$ representing a number drawn
from a known distribution \footnote{a known distribution is a
  distribution with known and computable moments} - in particular,
$c_i$ can be a real constant.
\item $U$ is the loop body and is a sequence of $m$ random updates, each of the form:
\begin{equation}\label{eq:ProbModel:prob_assignments}
x_i := a_i x_i + P_{i}(x_1,\dots x_{i-1}) \;[p_i]\; b_i x_i + Q_{i}(x_1,\dots x_{i-1}),
  \end{equation}
or, in case of a deterministic assignment, 
\begin{equation}\label{eq:ProbModel:det}
  x_i := a_i x_i + P_{i}(x_1,\dots x_{i-1}),
  \end{equation}
where $a_i, b_i \in \Real$ are constants and $P_{i},
Q_{i}\in\Real[x_1,\ldots,x_{i-1}]$ are polynomials over program
variables $x_1,\ldots,x_{i-1}$. Further,  $p_i\in[0,1]$
in~\eqref{eq:ProbModel:prob_assignments} is the
probability of updating $x_i$ to $a_i x_i + P_{i}(x_1,\dots x_{i-1})$,
whereas the probability to update $x_i$ to $b_i x_i + Q_{i}(x_1,\dots
x_{i-1})$ in~\eqref{eq:ProbModel:prob_assignments} is $1-p_i$.   
\end{itemize}

The coefficients $a_i$, $b_i$ and the coefficients of $P_i$ and
$Q_i$ in the variable assignments
\eqref{eq:ProbModel:prob_assignments}-\eqref{eq:ProbModel:det} of
\ProbModel{} loops can be drawn from a random distribution  as long as
the moments of this distribution are known and are independent from
program variables $x_1,\ldots,x_m$. Hence, the variable updates of
\ProbModel{} loop can involve coefficients drawn from Bernoulli,
uniform, normal, and other distributions. Moreover, \ProbModel{} support parametrised distributions, for
example one may have the random distribution ${\tt rand(d_1,d_2)}$ with
arbitrary $d_1,d_2\in\Real$ symbolic constants. Similarly, rather than only
considering concrete numeric values of $p_i$, 
the probabilities $p_i$ in the probabilistic updates~\eqref{eq:ProbModel:prob_assignments} of
\ProbModel{} loops can also be symbolic constants.
\begin{example}
The programs in Fig.~\ref{fig:pp1} are \ProbModel{},
using uniform distributions given by ${\tt rand()}$. Fig.~\ref{fig:pp1}(D) 
also uses normal distribution given by ${\tt gauss()}$.
Note that while the random distributions of Fig.~\ref{fig:pp1}(B,D) are
defined in terms of concrete constants, Fig.~\ref{fig:pp1}(A,C) have a
parametrised random distribution, defined in terms of $d\in\Real$. 
\end{example}
\vspace{-10pt}
\noindent\paragraph{\bf \ProbModel{} Loops and Moment-Based
  Recurrences.}
Let us now consider a \ProbModel{} program with $n\in\Nat$
denoting the loop iteration counter. 
We show that variable updates of \ProbModel{} programs yield special 
recurrences in $n$, called {\it moment-based recurrences}. For this, we consider program variables
$x_1,\ldots,x_m$ as sequences $x_1(n),\ldots,x_m(n)$ allowing us to
precisely describe relations between values of $x_i$ at different
loop iterations. Using this notation,
probabilistic updates~\eqref{eq:ProbModel:prob_assignments} over $x_i$
turn $x_i(n)$ into a random variable, yielding the 
%
relation (similarly, for deterministic
updates~\eqref{eq:ProbModel:det}):
\begin{equation*}\label{eq:probRec}
x_i(n+1) = a_i x_i(n) + P_i(x_1(n),\dots,x_{i-1}(n))
\:[p_i]\: b_i x_i(n) + Q_i(x_1(n),\dots,x_{i-1}(n)). 
\end{equation*}
The relation above could be treated as a recurrence equation over random variables
$x_i(n)$ provided the probabilistic behaviour depending on $p$ is
encoded (as an extension) into a recurrence equation. To analyse such  
probabilistic updates of \ProbModel{} loops, for each random variable
$x_i(n)$ we consider their expected values $E[x_i(n)]$ and
create new recurrence variables from expected 
values of monomials over original program variables (e.g. a new
variable $E[x_i\cdot x_j]$).
We refer to  these new recurrence variables  as \emph{E-variables}.
We note that any program variable yields an E-variable, but not every
E-variable corresponds to one single program variable as E-variables
are expected values of monomials over program variables. We now formulate
recurrence equations over E-variables rather than over program
variables, 
yielding {\it moment-based recurrences}. 

\begin{definition}[Moment-Based Recurrences]\label{def:MomentRec}
  Let $x(n)$ be a sequence of random variables. 
  A \emph{moment-based recurrence} for $x$ is a recurrence over
  E-variable $E[x]$: 
\[
E[x(n+r)]=R(E[x(n)], E[x(n+1)],\dots,E[x(n+r-1)],n)\quad (n\in \Nat),
\]
for some function $R:\Real^{r+1}\rightarrow\Real$, where $r\in\Nat$ is
the \emph{order} of the moment-based recurrence. 
\end{definition}

\noindent Similarly to~\cite{McIverM05}, 
note  that variable updates $x_i := f_1(x_i) \:[p_i]\: f_2(x_i)$
yield the relation: 
\begin{equation}\label{removing[]}
  \begin{array}{lcl}
	E[x_i(n+1)]
					&=& E\big[p_i\cdot f_1(x_i(n))+ (1-p_i)\cdot f_2(x_i(n))\big]\\
					&=& p_i\cdot E\big[f_1(x_i(n))\big] +
                                            (1-p_i)\cdot E\big[f_2(x_i(n))\big]
  \end{array}                                          
\end{equation}
Thanks to this relation,  probabilistic
updates~\eqref{eq:ProbModel:prob_assignments}
are rewritten into the moment-based recurrence equation:
\begin{equation}\label{eq:MomentRec}
  \begin{array}{lcl}
E[x_i(n+1)]
		&= & p_i\cdot  E\big[a_i x_i(n) + P_i(x_1(n),\dots,x_{i-1}(n))\big]\\
 		&& +(1-p_i)\cdot E\big[b_i x_i(n) +
           Q_i(x_1(n),\dots,x_{i-1}(n))\big].
\end{array}
\end{equation}
In particular, we have $E[x_i(n+1)] = p_i\cdot  E[a_i x_i(n) +
P_i(x_1(n),\dots,x_{i-1}(n))] $ for the deterministic assignments
of~\eqref{eq:ProbModel:det} (that is, $p_i=1$ in~\eqref{eq:ProbModel:det}).

By using properties of expected
values of expressions $expr_1,expr_2$ over random variables, we obtain
the following simplification rules:
\begin{equation}\label{eq:Evar:simplifRules}
  \begin{array}{lcl}
E[expr_1 + expr_2] &\rightarrow& E[expr_1] + E[expr_2]\\
E[expr_1 \cdot expr_2] &\rightarrow & E[expr_1] \cdot E[expr_2],
                                      \ {\small \text{if\ }
                                      expr_1,expr_2\text{ are independent}}\\
E[c\cdot expr_1] &\rightarrow & c\cdot E[expr_1]\\
E[c] &\rightarrow & c\\
E[\mathcal{D} \cdot expr_1] &\rightarrow& E[\mathcal{D}]
    \cdot E[expr_1]
    \end{array}\hspace*{-1.5em}
\end{equation}
where $c\in\Real$ is a constant and $\mathcal{D}$ is a known
independent distribution.

\begin{example}
  The moment-based recurrences of the \ProbModel{} loop of
  Fig.~\ref{fig:pp1}(A) are:
\begin{equation*}\label{eq:running:MBR}
\left\{ \begin{array}{lcl}
E[f(n+1)] &=& \frac{3}{4}E[1] + \frac{1}{4} E[0])\\
E[x(n+1)] &=& E\big[x(n) + f(n+1)\cdot rand(1-d,1+d)\big]\\
E[y(n+1)] &=& E\big[y(n) + f(n+1)\cdot rand(2-2d,2+2d)\big]\\
E[s(n+1)] &=& E\big[x(n+1) + y(n+1)\big]
        \end{array}\right.
    \end{equation*}
  By using the simplification rules~\eqref{eq:Evar:simplifRules} on
  the above recurrences,
  we obtain the following simplified moment-based recurrences
  of Fig.~\ref{fig:pp1}(A):
  \begin{equation}\label{eq:running:MBR_Simplified}
\left\{ \begin{array}{lcl}
E[f(n+1)] &=& \frac{3}{4}\\
E[x(n+1)] &=& E[x(n)] + E[f(n+1)] \cdot E[rand(1-d,1+d)]\\
E[y(n+1)] &=& E[y(n)] +E[f(n+1)] \cdot E[rand(2-2d,2+2d)]\\
E[s(n+1)] &=& E[x(n+1)] + E[y(n+1)]
\end{array}\right.
 \end{equation}
\end{example}

In Section~\ref{sec:algorithm} we show that \ProbModel{} loops can further be rewritten into a
system of C-finite recurrences over E-variables.


\vspace{-2mm}
\noindent\paragraph{\bf \ProbModel{} Loops  and Mutually Dependent
  Updates.}
Consider PP loops with mutually dependent affine updates: 
\begin{equation}\label{eq:ProbModel:probGen}
{\small x_i :=     \sum_{k=1}^m a_{i,k}x_k   \ + c_i   \;[p_i]\;
  \sum_{k=1}^m b_{i,k}x_k\ + d_i,}
\end{equation}
where $a_{i,k}, b_{i,k}, c_i, d_i \in\Real$ are constants. 
While such
assignments are not directly captured by updates~\eqref{eq:ProbModel:prob_assignments} of
\ProbModel{} loops, this is not a restriction of our work. Variable updates given 
by~\eqref{eq:ProbModel:probGen} yield mutually dependent C-finite
recurrences over E-variables. Using methods from~\cite{Kauers11}, this 
coupled system of C-finite recurrences can be rewritten into an
equivalent system of independent C-finite recurrences over E-variables,
yielding an independent system of moment-based recurrences over
which our invariant generation algorithm from Section~\ref{sec:algorithm} can be applied.
Hence probabilistic loops with
affine updates  are special cases of \ProbModel{} loops. 


%

\vspace{-2mm}
\noindent\paragraph{\bf Multi-Path \ProbModel{} Loops. }
While~\eqref{eq:ProbModel} defines \ProbModel{}
programs as single-path loops, the following class of multi-path loops
can naturally be modeled by \ProbModel{} programs: 
\begin{equation}\label{eq:ProbModelIf}
  I; \texttt{while(true)} \{\texttt{if } t \texttt{ then }U_1 \texttt{
    else } U_2\}, \qquad\qquad\text{where:}
\end{equation}

\noindent $I$ is as in~\eqref{eq:ProbModel}, $t$ is a boolean-valued random variable,  and $U_1$ and $U_2$ are respectively 
sequences of deterministic updates $x_i := a_i x_i + P_{i}(x_1,\dots
x_{i-1})$ and $x_i:=b_i x_i + Q_{i}(x_1,\dots x_{i-1})$ as
in~\eqref{eq:ProbModel:det}. 
PPs~\eqref{eq:ProbModelIf} can be rewritten to equivalent \ProbModel{}
loops, as follows. A pair of updates $x:=u_1[p]v_1$ from $U_1$ and
$x:=u_2[p]v_2$ from $U_2$  is  rewritten by the following sequence of updates: 
\begin{equation}\label{eq:ProbModelIfTranslated}
  \begin{array}{l}
    f:=1[p]0;\\
    g:=1[p]0;\\
    x:= t(u_1f + v_1(1-f)) + (1-t)(u_2g + v_2(1-g)) 
\end{array}
\end{equation}
with $f,g$ fresh program variables. The resulting program is
\ProbModel{} and we can thus compute moment-based invariants of
multi-path loops as in~\eqref{eq:ProbModelIf}. The programs {\sc
  Coupon, Random\_Walk\_2D} of Table~\ref{table:algorithm} are \ProbModel{} loops corresponding to such multi-path loops.

\section{Moment-Based Invariants of \ProbModel{} Loops}
\label{sec:algorithm}

Thanks to probabilistic updates, the values of program variables of \ProbModel{} loops after specific
number of loop iterations are not a priory determined. The 
value distributions $x_i(n)$ of program variables $x_i$ are
therefore random variables. When analysing \ProbModel{} loops, and in
general probabilistic programs, one is therefore required to capture
relevant properties over expected values and higher
moments of the variables in order to precisely summarise the value distribution of
program variables.

\noindent\paragraph{\bf Moment-Based Invariants.}
We are interested in automatically generating so-called
{\it moment-based
  invariants} of \ProbModel{} loops. Moment-based invariants are properties over expected values and higher moments of
program variables such that these properties  hold at
arbitrary loop iterations (and hence are invariants). 

\begin{algorithm}[t]
\caption{Moment-Based Invariants of \ProbModel{}\label{algo:InvGen} Loops}
\hspace*{\algorithmicindent} \textbf{Input:} \ProbModel{} loop
$\mathcal{P}$ as defined in~\eqref{eq:ProbModel}, with variables
$\{x_1,\dots,x_m\}$, and $k\geq 1$  \\
\hspace*{\algorithmicindent} \textbf{Output:} Set $MI$ of Moment-based
invariants of $\mathcal{P}$ over the $k$th moments of $\{x_1,\ldots, x_m\} $ \\
\hspace*{\algorithmicindent} \textbf{Assumptions:} $n\in\Nat$ is 
the loop counter of $\mathcal{P}$
\begin{algorithmic}[1]
  \State Extract the moment-based recurrence relations of
  $\mathcal{P}$\label{eq:algo:MBR1}, for $i=1,\ldots,m$:
  \[\begin{array}{lcl}
E[x_i(n+1)]
		&= & p_i\cdot  E\big[a_i x_i(n) + P_i(x_1(n),\dots,x_{i-1}(n))\big]\\
 		&& +(1-p_i)\cdot E\big[b_i x_i(n) +
           Q_i(x_1(n),\dots,x_{i-1}(n))\big].
      
    \end{array}\]
  \State $MBRecs=\{E[x_i(n+1)]~\mid~ i=1,\ldots,m\}$\label{eq:algo:MBR2}\Comment{initial set of
    moment-based recurrences}
 \State
 $S:=\{x_1^k,\ldots,x_m^k\}$\label{eq:algo:monomials}\Comment{initial
   set of monomials of E-variables}
 \Statex \hfill{as $\Mom{k}{0}{x_i(n)}=E[x_i(n)^k]$}
   \While{$S\not=\emptyset$}\label{eq:algo:loop_start}
        \State $M :=  \prod_{i=1}^{m}x_i^{\alpha_i} \in S$, where\label{algo:M}
        $\alpha_i\in\Nat$
        \State  $S := S\setminus\{M\}$
        
        \State $M' = M[x_i^{\alpha_i}\leftarrow upd_i]$,~ for each $i=m,\ldots,1$\label{algo:rewriteM}
        \Comment{replace each $x_i^{\alpha_i}$ in $M$ with $upd_i$}
        \Statex
        \Statex  \qquad\qquad where $upd_i$ denotes:
        \Statex \qquad\qquad$p_i\cdot \big(a_i x_i + P_{i}(x_1,\dots
        x_{i-1})\big)^{\alpha_i} + (1-p_i)\cdot\big( b_i x_i +
        Q_{i}(x_1,\dots x_{i-1})\big)^{\alpha_i}$
        \Statex
        \State Rewrite $M'$ as $M'=\sum
        N_j$ for monomials $N_j$ over $x_1,\ldots,x_m$\qquad\label{algo:MsumN}
        \State Simplify the moment-based recurrence $E[M(n+1)] = E[\sum
        N_j]$ using the \label{algo:simplifyM}
        rules~\eqref{eq:Evar:simplifRules} 
        \Statex \Comment{$M(n+1)$ denotes $\prod_{i=1}^{m}x_i(n+1)^{\alpha_i}$}
        \State $MBRecs=MBRecs\cup\{E[M(n+1)]\}$\label{algo:addMtoRecs}
        \Statex\Comment{add $E[M(n+1)]$ to the set of moment-based recurrences}
        \For{each monomial $N_j$ in $M$}
              \If{$E[N_j]\nin MBRecs$}\Comment{there is no
                moment-based recurrence for $N_j$} 
              \State $S=S\cup\{N_j\}$ \Comment{add $N_j$~to~$S$}\label{algo:Nj}
              \EndIf
        \EndFor
        
        \EndWhile
        \State {\bf end while}\label{eq:algo:loop_end}
  \State Solve the system of moment-based recurrences $MBRecs$\label{eq:algo:recSolve}
  \State $MI=\{E[x_i(n)^k]-CF_i(k,n)=0 \ \mid \ i=1,\ldots m\}$\label{algo:MI}
  \Statex\Comment{$CF_i(k,n)$ is the closed
    form solution of $E[x_i^k]$}
  \State \underline{\bf return} the set $MI$ of moment based invariants of
  $\mathcal{P}$ for the $k$th moments of $x_1,\ldots,x_m$
\end{algorithmic}
\end{algorithm}

\noindent\paragraph{\bf Automated Generation of Moment-Based
  Invariants of \ProbModel{} Loops.}
%
Our method for generating moment-based invariants of \ProbModel{}
loops is summarized in Algorithm~\ref{algo:InvGen}.
Algorithm~\ref{algo:InvGen} takes as input a \ProbModel{} loop
$\mathcal{P}$  and a natural number $k\geq 1$ and returns 
{\it moment-based invariants over the $k$th moments} of the program
variables $\{x_1,\ldots,x_m\}$. We denote by $n$ the loop
counter of $\mathcal{P}$.

\begin{theorem}\label{thm:line7Alg}
Higher-order moments of variables in \ProbModel{} loops can be modeled by C-finite recurrences over E-variables.
\end{theorem}
\begin{proof}[Sketch]
We want to show that $E[x_i^{\alpha_i}]$ can be expressed using recurrence equation. 
The idea is to express $x_i^{\alpha_i}(n+1)$ in terms of value of $x_i$ at $n$-th iteration. 
Value of $x_i(n+1)$ is $a_i x_i(n) + P_{i}(x_1(n+1),\dots x_{i-1}(n+1)$ with probability $p$ and $b_i x_i(n) + Q_{i}(x_1(n+1),\dots x_{i-1}(n+1)$ with probability $(1-p)$. From here we can derive that
$E[x_i^{\alpha_i}(n+1)] = E[p_i\cdot \big(
a_i x_i + P_{i}(x_1,\dots x_{i-1})\big)^{\alpha_i} + 
(1-p_i)\cdot\big( b_i x_i + Q_{i}(x_1,\dots x_{i-1})\big)^{\alpha_i}]$.
For arbitrary monomial $M = \prod x_i^{\alpha_i}(n+1)$ we can express $E[M]$ by 
substituting each $x_i^{\alpha_i}(n+1)$ as above. 
This process is captured by line~\ref{algo:rewriteM} of Alg.~\ref{algo:InvGen}.
The new equations can be further simplified using properties of expected values 
and the simplification rules~\eqref{eq:Evar:simplifRules}  
to give recurrence equations over E-variables.


\end{proof}

We now describe Algorithm~\ref{algo:InvGen}. Our algorithm first rewrites
$\mathcal{P}$ into a set $MBRecs$ of moment-based recurrences, as described
in Section~\ref{sec:programmingmodel}. That is, program variables $x_i$ are
turned into random variables $x_i(n)$ and variable updates over $x_i$
become moment-based
recurrences over E-variables by using the relation
of~\eqref{removing[]} (lines~\ref{eq:algo:MBR1}-\ref{eq:algo:MBR2}) of
Alg.~\ref{algo:InvGen}).

The algorithm next proceeds with computing the moment-based
recurrences of the $k$th moments of $x_1,\ldots,x_m$.
Recall that the $k$th moment of $x_i$ is given by:
$$\Mom{k}{0}{x_i(n)}=E[x_i(n)^k].$$
Hence, the set $S$ of monomials yielding E-variables for which
moment-based recurrences need to be solved is initialized to
$\{x_1^k,\ldots,x_m^k\}$ (line~\ref{eq:algo:monomials} of
Alg.~\ref{algo:InvGen}). Note that by considering the resulting E-variables
$E[x_i^k]$ and solving the moment-based recurrences of $E[x_i^k]$,
we derive closed forms of the $k$th moments of $\{x_1,\ldots,x_m\}$
(line~\ref{algo:MI} of Alg.~\ref{algo:InvGen}). To this end,
Algorithm~\ref{algo:InvGen} recursively computes the moment-based
recurrences of every E-variable arising from the moment-based
recurrences of $E[x_i^k]$
(lines~\ref{eq:algo:loop_start}-\ref{eq:algo:loop_end} of Alg.~\ref{algo:InvGen}), thus ultimately computing closed forms for
$E[x_i^k]$.  One can then use transformations described in
Proposition~\ref{prop:transformation} to compute closed forms for
other moments, such as variance and covariance.
In more detail, 
\begin{itemize}
  \item for each monomial $M = \prod x_j^{\alpha_j}$ from $S$, we
  substitute $x_i^{\alpha_i}$ in $M$ by its probabilistic behaviour. That is, the 
  update of $x_i$ in the \ProbModel{} loop
  $\mathcal{P}$ is rewritten, according to \eqref{removing[]},
  into the sum of its two probabilistic updates, weighted by
  their respective probabilities
  (lines~\ref{algo:M}-\ref{algo:rewriteM} of Alg.~\ref{algo:InvGen}).
  Rewriting in line~\ref{algo:rewriteM} of Alg.~\ref{algo:InvGen} 
  represents the most non-trivial step in our algorithm, 
  combining non-deterministic nature of our program with polynomial properties. 
 The resulting polynomial~$M'$ from~$M$ is then reordered to be expressed as a sum of
  new monomials  $N_j$ (line~\ref{algo:MsumN} of
  Alg.~\ref{algo:InvGen}); such a sum always exists as $M'$ involves
  only addition and multiplication over $x_1,\ldots,x_m$ (recall that
  $P_i$ and $Q_i$ are polynomials over $x_1,\ldots,x_m$). 
  \item By applying the simplification
    rules\eqref{eq:Evar:simplifRules} of E-variables over the
    moment-based recurrence of  $E[\sum N_j]$, the 
    recurrence of $E[M(n+1)]$ is obtained and added to the set $MBRecs$. Here, $M(n+1)$ denotes $\prod_{i=1}^{m}x_i(n+1)^{\alpha_i}$. As the recurrence of~$E[M(n+1)]$
    depends on $E[N_j]$, moment-based recurrences of $E[N_j]$ need
    also be computed and hence $S$ is enlarged by $N_j$
    (lines~\ref{algo:simplifyM}-\ref{algo:Nj} of Alg.~\ref{algo:InvGen}).
  \end{itemize}
  As a result, the set $MBRecs$ of moment-based recurrences of E-variables corresponding
  to $S$ are obtained. These recurrences are C-finite expressions over
  E-variables (see correctness argument of Theorem~\ref{thm:termination}) and hence
  their closed form solutions exist. In particular, the closed forms $CF_i(k,n)$ of
  $E[x_i(n)^k]$ is derived, turning 
  $E[x_i(n)^k]-CF_i(k,n)=0$ into a inductive property that holds at
  arbitrary loop iterations and is hence a moment-based invariant of
  $\mathcal{P}$ over the $k$th moment of $x_i$ (line~\ref{algo:MI} of
  Alg.~\ref{algo:InvGen}).

\begin{theorem}[Soundness]
Consider a \ProbModel{} loop $\mathcal{P}$ with program variables
$x_1,\ldots, x_m$ and let $k$ be a non-negative integer with 
$k\geq 1$. Algorithm~\ref{algo:InvGen} generates moment-based
invariants  of $\mathcal{P}$ over the $k$th moments of $x_1,\ldots,x_m$.
\end{theorem}

Note when $k=1$, Algorithm~\ref{algo:InvGen} computes the
moment-based invariants as
invariant relations over the closed form solutions of expected values
of $x_1,\ldots,x_m$. In this case, our moment-based invariants are
quantitative invariants as in~\cite{Katoen2010}. 

  \begin{example}
    We illustrate Algorithm~\ref{algo:InvGen} for computing the second
    moments (i.e. $k=2$) of the \ProbModel{} loop of
    Fig.~\ref{fig:pp1}(A). 
%
    Our algorithm initializes $MBRecs=\{E[f(n+1)], E[x(n+1)], E[y(n+1)],
    E[s(n+1)]\}$ and $S=\{f^2, x^2, y^2, s^2\}$.

We next (arbitrarily) choose $M$ to
be the monomial $f^2$ from $S$. Thus, $S=\{x^2,y^2,s^2\}$. Using the
probabilistic update of $f$, we replace $f^2$ by $\frac{3}{4}\cdot1^2 +
(1-\frac{3}{4})\cdot^2$, that is by $\frac{3}{4}$. As a result, $MBRecs=MBRecs\cup\{E[f(n+1)^2]=\frac{3}{4}\}$ and $S$ remains unchanged. 

We next choose $M$ to be $x^2$ and set
$S=\{y^2,s^2\}$. 
We replace $x^2$ by its randomised
behaviour, 
yielding $E[M(n+1)]=E[x(n+1)^2] = E[\big(x(n)+f(n+1)\cdot {\tt
  rand(1-d,1+d)}\big)^2]$.  By the simplification rules~\eqref{eq:Evar:simplifRules}
over E-variables, we obtain:
\begin{equation}\label{ex:xn}
  E[x(n+1)^2] =  E[x(n)^2] + 2\cdot E[x(n)]\cdot E[f(n+1)] +
  E[f(n+1)^2]\cdot \frac{1}{3}(d^2 + 3),
  \end{equation}
as $f(n+1)$ is independent from $x(n)$ and $E[{\tt
  rand(1-d,1+d)}^2]=\frac{1}{3}(d^2+3).$
We add the recurrence~\eqref{ex:xn} to
$MBRecs$ and keep $S$ unchanged  as the E-variables $E[x(n)], E[f(n+1)], E[f(n+1)^2]$
 have their recurrences
 already in $MBRecs$.

 We next set $M$ to $y^2$ and change $S=\{s^2\}$. Similarly to
 $E[x(n+1)^2]$, we get: 
 \begin{equation}\label{ex:yn}
  E[y(n+1)^2]=E[y(n)^2] + 4\cdot E[y(n)]\cdot E[f(n+1)] +
   E[f(n+1)^2]\cdot \frac{4}{3}(d^2 + 3), 
 \end{equation}
 by using that $f(n+1)$ is independent from $y(n)$ and $E[{\tt
   rand(2-2d,2+2d)}^2] = \frac{4}{3}(d^2 + 3)$. We add the
 recurrence~\eqref{ex:yn} to $MBRecs$ and keep $S$ unchanged.

 We set $M$ to $s^2$, yielding $S=\emptyset$. We extend $MBRecs$ with the recurrence: 
 \begin{equation*}\label{ex:sn}
   E[s(n+1)^2] = E[\big(x(n+1) + y(n+1)\big)^2] = E[x(n+1)^2] + 2 E[(xy)(n+1)] + E[y(n+1)^2] 
  \end{equation*}
 and add $xy$ to
 $S$. 
 We therefore consider $M$ to be $xy$ and set $S=\emptyset$. We
 obtain:
{\small \begin{equation*}
E[(xy)(n+1)] =  E[(xy)(n)] + 2\cdot E[x(n)]\cdot E[f(n+1)  +
E[y(n)]\cdot E[f(n+1)] + 2\cdot E[f(n+1)^2],
\end{equation*}}
by using that $E[{\tt rand(1-d,1+d)}]=1$ and $E[{\tt
  rand(2-2d,2+2d)}]=2$.
We add the recurrence of $E[(xy)(n+1)]$ to $MBRecs$ and keep
$S=\emptyset$.

As a result, we proceed to solve the moment-based recurrences of 
$MBRecs$. We focus first on the recurrences over expected values:
\begin{equation*}
  \begin{array}{lclcl}
    E[f(n+1)] &=& \frac{3}{4}\\
    E[x(n+1)] &=& E[x(n)]+E[f(n+1)\cdot{\tt rand(1-d,1+d)}] &= &  E[x(n)]+ \frac{3}{4}\\
    E[y(n+1)] &=& E[y(n)]+E[f(n+1)\cdot{\tt rand(2-2d,2+2d)}] &= &  E[x(n)]+2 \cdot\frac{3}{4}\\
    E[s(n+1)] &=& E[x(n+1)]+E[y(n+1)]
  \end{array}
\end{equation*}
Note that the above recurrences are C-finite recurrences over
E-variables.
For computing closed forms, we respectively substitute
$E[f(n+1)$ by its closed form in   $E[y(n+1)]$ and $E[x(n+1)]$, yielding
closed forms for $E[y(n+1)]$ and $E[x(n+1)]$, and hence for
$E[s(n+1)]$.
By also using the the initial values of Fig.~\ref{fig:pp1}, we derive the
closed forms: 
\begin{equation*}
  \begin{array}{lclclcl}
  E[f(n)] &=& \frac{3}{4}& \qquad  & E[s(n)]&=&\frac{9}{4}n\\
  E[x(n)] &=&\frac{3}{4}n-1 & \qquad&  E[y(n)] &=&\frac{3}{2}n+1
  \end{array}
\end{equation*}
We next similarly derive the closed forms for
higher-order and mixed moments:
\begin{equation*}
  \begin{array}{lclclcl}
  E[f(n)^2] &=& \frac{3}{4}& \qquad  & E[s(n)^2]&=&\frac{81}{16}n^2   +  \frac{20d^2+27}{16}n \\
  E[x(n)^2] &=&\frac{9}{16}n^2    +    \frac{4d^2-21}{16}n   +   1   &
                                                                       \qquad&  E[y(n)^2] &=&\frac{9}{4}n^2    +    \frac{4d^2+15}{4}n   +   1\\
     E[(xy)(n)] &=& \frac{9}{8}n^2    -    \frac{3}{8}n   -   1
  \end{array}
\end{equation*}
yielding hence 
the moment-based invariants over the second
moments of variables of Fig.~\ref{fig:pp1}. Using
Proposition~\ref{prop:transformation} and
Definition~\ref{def:common_Moments}, we derive the 
variance of $s(n)$ as $Var(s(n)) = \frac{20d^2+27}{16}n.$
\qed
\end{example}

Let us finally note that
the termination of Algorithm~\ref{algo:InvGen} depends whether for every monomial
        $M$ (from the set $S$, line~\ref{eq:algo:loop_start} of Alg.~\ref{algo:InvGen})
        the moment-based recurrence equation over the corresponding
        E-variable $E[M(n+1)]$ can be computed as C-finite
        expression over E-variables. To prove this, one can use
transfinite induction over monomials
. That is,  we order monomials and
show that the recurrence of $E[M(n+1)]$ depends only on
smaller monomials, for which we can compute C-finite closed form expressions. Thus
we have an inhomogeneous C-finite recurrence relation for $E[M(n+1)]$,
yielding a C-finite closed form expression. 
        
\begin{theorem}[Termination]
\label{thm:termination}
For any non-negative integer $k$ with $k\geq 1$ and any \ProbModel{} loop $\mathcal{P}$ with program variables
$x_1,\ldots, x_m$, Algorithm~\ref{algo:InvGen} terminates.
\\
Moreover,  Algorithm~\ref{algo:InvGen} terminates in at most
$\mathcal{O}(  k^m \cdot d_{m}^{m-1} \cdot d_{m-1}^{m-2} \ldots \cdot d_2^1)$ steps, where $d_i=max\{deg(P_i),
deg(Q_i), 1\}$ with $deg(P_i),
deg(Q_i)$ denoting the degree of polynomials $P_i$ and $Q_i$ of the
variable updates~\eqref{eq:ProbModel:prob_assignments}.
\end{theorem}

\begin{proof}
  We associate every monomial with an ordinal number as follows:
  $$ x_k^{\alpha_k} \cdot  x_{k-1}^{\alpha_{k-1}} \dots x_1^{\alpha_1}    \xrightarrow{\sigma}  \omega^k\cdot \alpha_k + \omega^{k-1}\cdot \alpha_{k-1} \dots + \alpha_{1},$$
		and order monomials $M,N$ such that $M > N$ if $\sigma(M) > \sigma(N)$. 
	Algorithm~\ref{algo:InvGen} terminates if for every monomial
        $M$ (from the set $S$, line~\ref{eq:algo:loop_start} of Alg.~\ref{algo:InvGen})
        the moment-based recurrence equation over the corresponding
        E-variable  $E[M(n+1)]$ can be computed as C-finite
        expression over E-variables. We will show that this is indeed the case by transfinite induction over monomials. 
	
	Let $M = \prod_{k = 1}^{K} x_k^{\alpha_k}$ be a monomial and assume that every smaller monomial has a closed form solution in form of a C-finite expression. 
	
	Let 
	\begin{equation}
		x_i^{\alpha_i} := \big(c_ix_i + P_i(x_1,\cdots x_{i-1})\big)^{\alpha_i}
	\end{equation}
	be the updates of our variables after removing the probabilistic choice as in line~\ref{algo:M} of the algorithm. Then recurrence for $M$ is
	\begin{align}\label{eq:simp_monomial}
		E[M(n+1)]
					&= E\Big[ \prod_{i=1}^{K}  \Big( p_i\cdot \big(a_i x_i + P_{i}(x_1,\dots x_{i-1})\big)^{\alpha_i} \nonumber\\
        & \quad + (1-p_i)\cdot\big( b_i x_i + Q_{i}(x_1,\dots x_{i-1})\big)^{\alpha_i}\Big)(n)\Big]\nonumber\\
					&= E[M(n)] + \sum_{j=1}^J b_j\cdot E\big[N_j(n)\big]
	\end{align}
	for some $J$, constants $b_i$ and monomials $N_1,\dots, N_J$
        all different than $M$. By Lemma~\ref{lemma:reduction}, we have an inhomogeneous C-finite recurrence relation 
	$E[M(n+1)] = E[M(n)] + \gamma $, for some C-finite expression
        $\gamma$. Hence, the closed form of $E[M(n+1)]$ exists and is
        a C-finite expression.
	\qed
\end{proof}

We finally prove our auxiliary lemma. 
\begin{lemma}\label{lemma:reduction}
		$M>N_j$ for all $j\le J$ in~\eqref{eq:simp_monomial}.
	\end{lemma}
\begin{proof}
	Let $M = \prod_{k = 1}^{K} x_k^{\alpha_k}$ and have $N_j = \prod_{k = 1}^{K} x_k^{\beta_k}$ coming from 
	\begin{equation}\label{eq:lemma}
		\prod_{i=1}^{K}  \big(   c_ix_i + P_i(x_1,\cdots x_{i-1})   \big)^{\alpha_i}.
	\end{equation}
	
	Assume $M \le N_j$, i.e. $\omega^K\cdot \alpha_K + \dots +
        \alpha_{1} \le \omega^K\cdot \beta_K + \dots + \beta_{1} $, so
        we have $\alpha_K \le \beta_K$. Note that in~\eqref{eq:lemma} $x_K$ only appears in factor $c_Kx_K + P_K(x_1,\dots x_{K-1})$.
Considering the multiplicity, we get at most $\alpha_K$th power of $x_K$, hence 	$\alpha_K \ge \beta_K$. Thus $\alpha_K = \beta_K$

So for $M \le N_j$ we need $N_j$ from $(c_Kx_K)^{\alpha_K} \cdot \prod_{i=1}^{K-1}  \big(   c_ix_i + P_i(x_1,\cdots x_{i-1})   \big)^{\alpha_i}$.

Proceeding similarly for $x_{K-1}, x_{K-2}, \cdots$ we get that for each $k\le K$ we have $\alpha_k=\beta_k$,  which contradicts the assumption, thus $M > N_j$ as needed.

Regarding the termination, let's look at what monomials can possibly be added to $S$. Let $M = \prod x_i^{\alpha_i} \in S$. Based on the algorithm and above it is clear that in case $i=m$ we have $\alpha_m \le k$. For any $i<m$ the maximum value of $\alpha_i$ is  $\alpha_{i+1}\cdot d_{i+1}$. Hence we have $\alpha_i \le k\cdot \prod_{j = i+1}^m d_j$. thus we can count all possible monomials, hence the upper bound on the algorithm time complexity, as product of theses upper bounds. This yields $k^m \cdot d_{m}^{m-1} \cdot d_{m-1}^{m-2} \ldots \cdot d_2^1$ as claimed.
\qed
\end{proof}


\section{Implementation and Experiments}
\label{sec:implementation}


We implemented our work in the Julia language, using  
\Aligator{}\cite{Aligator18} for handling and solving recurrences. 
We evaluated our work on several challenging probabilistic programs with parametrised
distributions, symbolic probabilities and/or both discrete and continuous random variables. 
All our experiments were run on MacBook Pro 2017 with 2.3 GHz Intel Core i5 and 8GB RAM.
Our implementation and benchmarks are available at: \url{github.com/miroslav21/aligator}.

%
 \noindent
\paragraph*{\bf Benchmarks.} 
We evaluated our work on 13 probabilistic programs, as follows. 
We used 7 programs from
works~\cite{Chen2015,Katoen2010,Chakarov2014,Feng2017,Kura19} on
invariant generation. These examples are given in lines 1-7 of
Table~\ref{table:algorithm}; we note though that \textsc{Binomial("$p$")} represents our generalisation of a binomial
distribution example taken from~\cite{Chen2015, Feng2017, Katoen2010} 
 to a probabilistic program with parametrised probability $p$. 
We further crafted 6 examples of our own, illustrating the 
distinctive features of our work. These examples are listed in lines
8-13 of Table~\ref{table:algorithm}: lines 8-11 correspond to the
examples of Fig.~\ref{fig:pp1};  line~12 of Table~\ref{table:algorithm}
shows a variation of Fig.~\ref{fig:pp1}, with a parametrized
distribution $p$; line~13 corresponds to a non-linear \ProbModel{}
loop computing squares. 
All our benchmarks are also available at the aforementioned url.

\begin{multicols}{2}

\subsection{\textsc{Coupon}}
Probabilistic model of Coupon Collector's program for two coupons, taken from \cite{Kura19}. 
\begin{lstlisting}
f := 0
c := 0
d := 0
while true:
   f := 1 [1/2] 0
   c := 1 - f + c*f
   d := d + f - d*f
\end{lstlisting}

\subsection{\textsc{Coupon4}}
Probabilistic model of Coupon Collector's program for four coupons, taken from \cite{Kura19}. 
\begin{lstlisting}
f := 0
g := 0
a := 0
b := 0
c := 0
d := 0
while true:
   f := 1 [1/2] 0
   g := 1 [1/2] 0
   a := a + (1-a)*f*g
   b := b + (1-b)*f*(1-g)
   c := c + (1-c)*(1-f)*g
   d := d + (1-d)*(1-f)*(1-g)
 \end{lstlisting}

\subsection{\textsc{Random\_walk\_1D\_cts}}
A variation of random walk in one dimension with assignments from continuous distributions taken from \cite{Kura19}.
\begin{lstlisting}
v := 0
x := 0
while true:
    v := u(0,1)
    x := x + v [7/10] x - v
\end{lstlisting}

\subsection{\textsc{Sum\_rnd\_series}}
A program modeling \textit{Sum of Random Series} game taken from \cite{Chen2015}.
\begin{lstlisting}
n := 0
x := 0
while true:
    n := n + 1
    x := x + n [1/2] x
\end{lstlisting}

\subsection{\textsc{Product\_dep\_var}}
A program modeling \textit{Product of Dependent Random Variables} game taken from \cite{Chen2015}.
\begin{lstlisting}
f := 0
x := 0
y := 0
p := 0
while true:
    f := 0 [1/2] 1
    x := x + f
    y := y + 1 - f
    p := x*y
\end{lstlisting}

\subsection{\textsc{Random\_walk\_2D}}
A variation of random walk in two dimension as in \cite{Chakarov2014, Kura19}.
Each direction is chosen with equal probability.
\begin{lstlisting}
h := 0
x := 0
y := 0
while true:
    h := 1 [1/2] 0
    x := x-h [1/2] x +h
    y := y+(1-h) [1/2] y-(1-h)
\end{lstlisting}

\subsection{\textsc{Binomial}}
Another classic example, modeling binomial distribution. Appeared also in \cite{Chen2015, Feng2017, Katoen2010}. However, we consider the program to be parametric,  computing invariants for arbitrary values of $p$.
\begin{lstlisting}
x := 0
while true:
    x := x + 1 [p] x
\end{lstlisting}

\subsection{\textsc{StutteringA}}
Program \ref{fig:pp1}(A) from Introduction.

\subsection{\textsc{StutteringB}}
Program \ref{fig:pp1}(B) from Introduction.

\subsection{\textsc{StutteringC}}
Program \ref{fig:pp1}(C) from Introduction.

\subsection{\textsc{StutteringD}}
Program \ref{fig:pp1}(D) from Introduction.

\subsection{\textsc{StutteringP}}
A variation of program \ref{fig:pp1}(A) from Introduction with $d=1$, parametrized w.r.t. $p$.
\begin{lstlisting}
f := 0
x := -1
y := 1
s := 0
while true:
    f := 1 [p] 0
    x := x + f*u(0,2)
    y := y + f*u(0,4)
    s := x + y
\end{lstlisting}

\subsection{\textsc{Square}}
Our own program with polynomial assignments.
\begin{lstlisting}
x := 0; y := 1
while true:
    x := x+2 [1/2] x
    y := x^2
\end{lstlisting}

\end{multicols}

\noindent
\paragraph*{\bf Experimental Results with Moment-Based Invariants.}
Results of our evaluation are presented in Table~\ref{table:algorithm}. 
While Algorithm~\ref{algo:InvGen} can compute invariants over arbitrary $k$th
higher-order moments, due to lack of space and readability, 
Table~\ref{table:algorithm} lists only our moment-based invariants up
to the third moment (i.e. $k\leq 3$), that is for expected values, second- and third-order moments. 
The first column of Table~\ref{table:algorithm} lists the benchmark
name, whereas the second column gives the degree of the moments
(i.e. $k=1, 2, 3$) for which we compute invariants. 
The third column reports the timings (in seconds) our implementation 
needed to derive invariants.
The last column shows our moment-based invariants; for readability, 
we decided to omit intermediary invariants 
(up to $30$ for some programs) and only show the most relevant
invariants. 

We could not perform a fair practical comparison with other existing
methods:  to the best of our knowledge, existing works, such
as~\cite{Katoen2010,GretzKM13,Barthe2016,Kura19}, require user guidance/templates/hints. Further, existing techniques do not support symbolic
probabilities and/or 
parametrised distributions - which are, for example, required in the
analysis of programs {\sc StutteringA, StutteringC, StutteringP} of
Table~\ref{table:algorithm}. 
We also note that examples {\sc Coupon, StutteringC, StutteringP} involve
non-linear probabilistic updates hindering automation in existing
methods, while such updates can naturally be encoded as moment-based recurrences
in our framework.
We finally note that while second-order moments are computed only
by~\cite{Kura19}, but  with the
help of user-provided templates, no existing approaches
compute moments 
for $k\geq 3$. Our experiments show that inferring third-order moments
are in general not expensive; yet, for examples {\sc StutteringA, StutteringC, StutteringP}  with parametrized
distributions/probabilities more computation time is needed.

\begin{table}[]
\scriptsize
\begin{center}
\begin{tabular}{ | l | c | c | p{0.65\linewidth} | l | l | l | }
\hline
\textbf{Program} & \textbf{Moment}                                          & \textbf{Runtime ($s$)}          & \textbf{Computed Moment-Based Invariants                                                                                                                                                                                                                                                                                                                     }    \\ \hline\hline
\multirow{3}{*}{\textsc {Coupon}~\cite{Kura19}} &  1                                                 & 0.37    & $E[c(n)] = (2^n - 1)/(2^n)$                                                                                                                                                                                                                                                                                                                     \\ \cline{2-4}
& 2                                                 & 0.40    & $E[c^2(n)] = (2^n - 1)/(2^n)$                                                                                                                                                                                                                                                                                                                   \\ \cline{2-4}
                 & 3                                                 & 0.34    & $E[c^2(n)] = (2^n - 1)/(2^n)$                                                                                                                                                                                                                                                                                                                   \\ \hline\hline
\multirow{3}{*}{\textsc {Coupon4}~\cite{Kura19}} & 1                                              & 0.90    & $E[c(n)] = (4^n - 3^3) / (4^n)$                                                                                                                                                                                                                                                                                                                 \\ \cline{2-4}
& 2                                               & 1.1     & $E[c^2(n)] = (4^n - 3^3) / (4^n)$                                                                                                                                                                                                                                                                                                               \\\cline{2-4} 
                 & 3                                                & 1.3     & $E[c^3(n)] = (4^n - 3^3) / (4^n)$                                                                                                                                                                                                                                                                                                               \\ \hline\hline
  \multirow{3}{*}{\textsc{random\_walk\_1d\_cts}~\cite{Kura19} } & 1                                  & 0.12    & $E[x(n)] = n/5$                                                                                                                                                                                                                                                                                                                                 \\ \cline{2-4}
& 2                                  & 0.45    & $E[x^2(n)] = n^2/25 + 22 n/75$                                                                                                                                                                                                                                                                                                                  \\\cline{2-4} 
                 & 3                                   & 1.00    & $E[x^3(n)] = n^3/125 + n^2 22/125 - n 21/250$                                                                                                                                                                                                                                                                                                   \\ \hline\hline
  \multirow{3}{*}{\textsc{sum\_rnd\_series}~\cite{Chen2015}} & 1                                      & 0.31    & $E[x(n)] = n^2/4 + n/4$                                                                                                                                                                                                                                                                                                                         \\ \cline{2-4}
& 2                                      & 2.89    & $E[x^2(n)] = n^4/16 + 5 n^3/24 + 3 n^2/16 + n/24$                                                                                                                                                                                                                                                                                               \\ \cline{2-4}
                 & 3                                      & 17.7    & $E[x^3(n)] = n^6/64 + 7 n^5/64 + 13 n^4/64 + 9 n^3/64 + n^2/32$                                                                                                                                                                                                                                                                                 \\ \hline\hline
  \multirow{3}{*}{\textsc{product\_dep\_var}~\cite{Chen2015}} & 1                                     & 0.65    & $E[p(n)] =  n^2/4 - n/4$                                                                                                                                                                                                                                                                                                                        \\ \cline{2-4}
& 2                                     & 6.27    & $E[p^(n)] = n^4/16 - n^3/8 + 3 n^2/16 - n/8$                                                                                                                                                                                                                                                                                                    \\ \cline{2-4}
                 & 3                                    & 37.5    & $E[p^3(n)] =  n^6/64 - 3 n^5/64 + 9 n^4/64 - 21 n^3/64 + 15 n^2/32 - n/4$                                                                                                                                                                                                                                                                       \\ \hline\hline
\multirow{3}{*}{\textsc{random\_walk\_2d}~\cite{Chakarov2014, Kura19} } & 1                         & 0.07    & $E[x(n)] = 0$                                                                                                                                                                                                                                                                                                                                   \\\cline{2-4} 
& 2                          & 0.26    & $E[x^2(n)] = n/2$                                                                                                                                                                                                                                                                                                                               \\\cline{2-4}
& 3                          & 0.49    & $E[x^3(n)] = 0$                                                                                                                                                                                                                                                                                                                                 \\ \hline\hline
\multirow{3}{*}{\textsc{Binomial("$p$")}~\cite{Chen2015, Feng2017,Katoen2010} } & 1                 & 0.17    & $E[x(n)] = n p$                                                                                                                                                                                                                                                                                                                                 \\\cline{2-4}
& 2                  & 0.47    & $E[x^2(n)] = n^2 p^2 + n p (1-p)$                                                                                                                                                                                                                                                                                                               \\ \cline{2-4}
                 & 3                  & 1.6     & $E[x^3(n)] = n^3 p^3 - 3 n^2 p^3 + 3 n^2 p^2 + 2 n p^3 - 3 n p^2 + n p$                                                                                                                                                                                                                                                                         \\ \hline\hline
  
  \multirow{3}{*}{\textsc{StutteringA -- Fig.~\ref{fig:pp1}(A)}} & 1                                                          & 0.44     & $E[s(n)] = 9 n/4$                                                                                                                                                                                                                                                                                                                               \\\cline{2-4} 
& 2                                                         & 2.2
                                                                                                              & $E[s^2(n)] = 81 n^2/16 + (20 d^2 + 27)/16   n$        \\\cline{2-4}
  & 3 & 8.48 & $E[s^3(n)]=81d^2n^2/16 + 63d^2n/16 + 729n^3/64 + 9n^2(4d^2 - 9)/32 + 9n^2(4d^2 + 9)/16 + 567n^2/64 + 3n(-6d^2 - 21)/8 + 3n(6d^2 - 12)/16 + 243n/32$\\ \hline\hline
  \multirow{3}{*}{\textsc{StutteringB -- Fig.~\ref{fig:pp1}(B)}} & 1                                                               & 0.49     & $E[s(n)] = 9 n/4$                                                                                                                                                                                                                                                                                                                               \\ \cline{2-4}
& 2                                                              &
                                                                   2.03
                                                                                                              &
                                                                                                                $E[s^2(n)]
                                                                                                                =
                                                                                                                81
                                                                                                                n^2/16
                                                                                                                +
                                                                                                                347/16
                                                                                                                n
                                                                                                                +
                                                                                                                128/3$
  \\\cline{2-4}
                 & 3 & 7.43 & $E[s^3(n)]=729n^3/64 + 9369n^2/64 + 1359n/32$=\\ \hline\hline
\multirow{3}{*}{\textsc{StutteringC -- Fig.~\ref{fig:pp1}(C)}} &  1                                                          & 1.8     & $E[s(n)] = 3 n^3/8 + 3 n^2/8 - n$                                                                                                                                                                                                                                                                                                               \\ \cline{2-4}
& 2                                                        & 72.5     & $E[s^2(n)] = 9 n^6/64 + 3 n^5 (8 d^2 + 27)/160 + n^4 (8 d^4 + 84 d^2 - 90)/192 + n^3 (32 d^4 + 216 d^2 - 252)/288 + n^2 (8 d^4 + 44 d^2 + 61)/64 + n (80 d^4 + 324 d^2 - 9)/1440$                                                                                                                                                               \\  \cline{2-4}
                 & 3                                                        & 2144     & $E[s^3(n)] = 27 n^9/512 + 27 n^8 (16 d^2 + 39)/2560 + 3 n^7 (824 d^4 + 6444 d^2 + 1242)/17920 + n^6 (1900 d^4 + 3996 d^2 - 4365)/2560 + n^5 (2004 d^4 + 1704 d^2 - 54)/2560 + n^4 (-1900 d^4 - 7056 d^2 + 13446)/7680 + n^3 (-6948 d^4 - 12708 d^2 - 6969)/7680 + n^2 (-1900 d^4 - 3114 d^2 - 315)/3840 + n (-108 d^4 - 603 d^2 + 288)/6720$    \\ \hline\hline
  \multirow{3}{*}{\textsc{StutteringD -- Fig.~\ref{fig:pp1}(D)}} & 1
                                                                            & 1.92
                                                                                                              & $E[s(n)]=3n^3/8 + 3n^2/8 - n$
                                                                                                                \\\cline{2-4}
                 & 2 &46.3 & $E[s^2(n)]=9n^6/64 + 93n^5/32 + 1651n^4/96 + 2849n^3/72 + 2813n^2/64 + 5131n/288$ \\\cline{2-4}
    & 3 & 2076& $E[s^3(n)]= 27n^9/512 + 1593n^8/512 + 94587n^7/1792 + 545971n^6/2560 + 270117n^5/1280 - 58585n^4/768 - 132599n^3/512 - 536539n^2/3840 - 771n/140$\\\hline\hline
  \multirow{3}{*}{\textsc{StutteringP}} & 1
                                                                            & 0.28
                                                                                                              &$E[s(n)]= 3np$
                                                                                                                \\\cline{2-4}
                 & 2 &1.68 &$E[s^2(n)]= 11n^2p^2 + 3np(-2p + 1) + np(-p - 1) + 4np(-p + 2) - 1$ \\\cline{2-4}
                 & 3 & 6.05 &$E[s^3(n)]= 27n_1^3p^3 - 3n_1^2p^3 + 3n_1^2p^2(-6p + 3) + 12n_1^2p^2(-3p + 3) + 12n_1^2p^2(-2p + 3) + 3n_1p(4p^2 - 3p + 3) + 3n_1p(8p^2 - 12p + 9) + n_1p(p^2 - 3p(-p - 1) - 3p + 2)/2 + 2n_1p(2p^2 - 6p(-p + 2) - 6p + 13) + 6$ \\\hline\hline
  \multirow{3}{*}{\textsc{Square}} & 1 & 0.38 & $E[y(n)]=n^2 + n$
                                                                                                                \\\cline{2-4}
                 & 2 & 2.46 & $E[y^2(n)]=n^4 + 6*n^3 + 3*n^2 - 2*n$ \\\cline{2-4}
    & 3 & 8.70 & $E[y^3(n)]=n^6 + 15*n^5 + 45*n^4 - 15*n^3 - 30*n^2 + 16*n$ \\\hline\hline
  \end{tabular}
\caption{{\small Moment-based
  invariants of \ProbModel{}~loops, where $n$ is the loop counter.}}
\label{table:algorithm}
\end{center}
\end{table}

\section{Related Work}
\label{sec:related}


Despite the impressive recent advancements, probabilistic model 
checking tools~\cite{Baier2008} (e.g., PRISM~\cite{KwiatkowskaNP11}, 
STORM~\cite{DehnertJK017} and MRMC~\cite{KatoenZHHJ11}) are not 
able to handle programs with unbounded and real variables.  Model checking 
algorithms suffer from the state explosion problem and their performance 
in terms of time and memory consumption degrades as the number of 
reachable states to be considered increases. Furthermore, 
probabilistic model checking tools have no support for invariant
generation. 
Our approach, based on symbolic summation over probabilistic expressions, can instead analyse 
probabilistic programs with a potentially infinite number of reachable 
states. 

In~\cite{McIverM05}, one of the first deductive 
frameworks to reason about probabilistic programs was proposed by 
annotating probabilistic programs with 
real-valued 
expressions over the expected values of program variables. 
Of particular interest are the annotations as \emph{quantitative
  invariants}, summarising loop behaviors.
The setting of~\cite{McIverM05} considers probabilistic programs where the 
stochastic inputs are restricted to discrete distributions with 
finite support and can deal also with demonic non-deterministic 
choice.  Although our approach does not yet support demonic 
non-determinism, we are not restricted to discrete input distributions 
as long as we know their moments (e.g., the Gaussian distribution 
is characterised only by two moments: the mean and the
variance). Moreover, our work is not restricted to quantitative
invariants as invariants over expected values of program
variables. Rather, we generate moment-based invariants that precisely
capture invariant properties of higher-order and mixed moments of
program variables.  


Katoen et al. provided in~\cite{Katoen2010}  the first semi-automatic and complete 
method synthesising the linear quantitative invariants needed by~\cite{McIverM05}. 
The work of~\cite{Katoen2010}, implemented in PRINSYS~\cite{GretzKM13},  
consists in annotating a loop with a linear template invariants and uses a constraint solver 
to find the parameters for which the template yields an invariant. 
The works~\cite{Feng2017,Chen2015} also synthesize 
non-linear quantitative invariants. 

Another related line of research is given in~\cite{Barthe2016}, where
martingales are used to compute invariants of probabilistic programs. 
The martingales generated
by~\cite{Barthe2016} however heavily depend on the user-provided hints and
hence less generic hints yield less expressive/precise
invariants. Moreover,  
of~\cite{Barthe2016} mainly focuses on invariants over expected values
and it remains unclear which 
extensions of martingales need to be considered to compute
higher-order moments. 
The work of~\cite{Kura19} addresses such  generalizations  of
martingales for computing higher-order moments of program
variables, with the overall goal of approximating runtimes of
randomized programs.
The approach in~\cite{Kura19} is however again restricted to
user-provided templates. Unlike the works
of~\cite{Katoen2010,GretzKM13,Feng2017,Chen2015,Barthe2016,Kura19}, our work does not rely on a priori given
templates/hints,
but computes the most precise invariant expression over higher-order or
mixed moments of program variables. To do so, we use symbolic
summation to compute closed forms of higher-order moments. In addition, 
\ProbModel{} loops support  parametrized distributions and symbolic
probabilities, which is not the case of~\cite{Barthe2016,Kura19}. 

There are two orthogonal problems related to quantitative 
invariants generation: {program termination}~\cite{McIverMKK18,FuC19}  and 
{worst-case execution}~\cite{BouissouGPCS16,Karp94,pldi2019}. 
The first is to assess whether a probabilistic program terminates  
with probability 1 or if the expected time of termination is bounded.  
In principle, one can use our approach to solve 
this class of problems for \ProbModel{} loops, but this is not 
the focus of this paper. 
The second class of problems is related to finding 
bounds over the expected values. In~\cite{BouissouGPCS16} 
the authors consider bounds also over higher-order moments
 for a specific class of probabilistic programs with 
  {probabilistic affine assignments}. This approach 
  can handle also nonlinear terms using interval arithmetics 
  and fresh variables, at the price to produce very 
  conservative bounds. On the contrary our approach 
  supports natively probabilistic polynomial assignments 
  (in the form of \ProbModel{}  loops)  and provides a precise 
  symbolic expression over  higher-order moments. 


\vspace*{-.5em}
\section{Conclusion}
\label{sec:conclusion}
\vspace*{-.5em}

We introduced a novel approach for automatically
generating {\it moment-based invariants} of a subclass of
probabilistic programs (PPs), called \ProbModel{} loops, with
polynomial assignments over random variables and
parametrised distributions.  We combine
methods from symbolic summation and statistics to
derive invariants over higher-order
moments, such as expected values or variances, of
program variables.  To the best of our knowledge, our
approach is the first method computing higher-order moments
of PPs fully automatically and the first
to handle PPs with  parametrised distributions.

%
%
\bibliographystyle{splncs04}
\bibliography{mybib_compact}


\end{document}